\documentclass[twocolumn,amssymb,amsmath,nofootinbib]{revtex4-1}
\usepackage{mathrsfs}
\DeclareMathAlphabet{\mathpzc}{OT1}{pzc}{m}{it}

\usepackage[english]{babel}

\usepackage{amsmath}
\usepackage{amssymb}

\usepackage{mathrsfs}

\usepackage{cancel} 
\usepackage{paralist}
\usepackage{booktabs}
\usepackage{hyperref}
\usepackage{xspace}

\usepackage{graphicx,color,verbatim,acronym,epsfig,subfigure}
\usepackage{slashed,skak}
\usepackage{latexsym,amsfonts,amssymb,amsmath,makeidx,mathrsfs,multirow,lineno,bm,bbm}

\renewcommand{\exp}[1]{\ \mathrm{exp} \left[ #1 \right]}

\DeclareSymbolFont{bbold}{U}{bbold}{m}{n}
\DeclareSymbolFontAlphabet{\mathbbold}{bbold}

\newcommand{\GeV}{\ \mathrm{GeV}}

\begin{document}
\title{Probing the Electroweak Sphaleron with Gravitational Waves}
\author{Ruiyu Zhou $^{1}$}

\author{Ligong Bian $^{1}$}
\email{lgbycl@cqu.edu.cn}
\author{Huai-Ke Guo$^{2}$}
\email{ghk@ou.edu}

\affiliation{
	$^1$~Department of Physics, Chongqing University, Chongqing 401331, China\\
$^2$ Department of Physics and Astronomy, University of Oklahoma, Norman, OK 73019, USA
}
\date{\today}

\begin{abstract}

We present the relation between the sphaleron energy and the gravitational
wave signals from a first order electroweak phase transition. The crucial ingredient is the scaling law between the sphaleron energy at 
the temperature of the phase transition and that at zero temperature. We estimate the baryon number preservation criterion, and observe that for 
a sufficiently strong phase transition, it is possible to probe the electroweak sphaleron using measurements of future space-based gravitational wave 
detectors.

\end{abstract}

\graphicspath{{figure/}}

\maketitle
\baselineskip=16pt

\pagenumbering{arabic}

\vspace{1.0cm}

\section{Introduction}

The observation of gravitational wave signals from the Binary Black
hole merger by LIGO~\cite{Abbott:2016blz} and the approval of the space-based interferometer LISA by the European Space Agency~\cite{Seoane:2013qna} 
have raised increasing interest on the study of gravitational waves from the Electroweak phase transition (EWPT) in the early Universe. To account for the  baryon asymmetry of the Universe (BAU), the mechanism of electroweak baryogenesis (EWBG) requires a strongly first order electroweak phase transition (SFOEWPT). The SFOEWPT provides a non-equilibrium environment for baryon number generation~\cite{ Kuzmin:1985mm}(see~\cite{Morrissey:2012db,Mazumdar:2018dfl} for recent reviews on EWBG and cosmic phase transitions), fulfilling one of the three Sakharov conditions~\cite{Sakharov:1967dj}. 
The (B+L)-violating sphaleron process associated with the change of Chern-Simons numbers~\cite{Manton:1983nd,Klinkhamer:1984di} should be suppressed to avoid the washout of the baryon asymmetry inside the electroweak bubbles (electroweak broken phase) after the EWPT~\cite{Kuzmin:1985mm}.
Particularly, the sphaleron rate in the broken phase is proportional to a Boltzmann factor $\Gamma\propto \exp{-E_{sph}(T)/T}$~\cite{Kuzmin:1985mm,Rubakov:1996vz}, with here $E_{sph}(T)$ representing the sphaleron energy (energy barrier) of the 
saddle-point configuration of the electroweak theory~\cite{Klinkhamer:1984di}.
The requirement that the sphaleron process needs to be sufficiently quenched constrains the possible patterns of EWPT and thus the generated gravitational waves since $E_{sph}(T)$ is highly correlated with the Higgs VEV at finite temperature as will be explored in the following. We therefore propose to probe the sphaleron process at the finite temperature through the measurement of the gravitational wave signals.

We start by studying the relation between the sphaleron energy and the strength of the EWPT.
Requiring the sphaleron rate in the broken phase to be lower than the Hubble expansion rate results in 
the baryon number preservation criterion (BNPC)~\cite{Gan:2017mcv}\footnote{Here, we note that the exact settling down of this relation requires 
lattice simulation of the sphaleron rate. See Ref.~\cite{DOnofrio:2014rug} for a recent study in the SM.},
\begin{align}\label{eq:Esph_bound}
	\frac{E_{\rm sph}(T)}{T} > (35.9-42.8) + 7 \ln \frac{v(T)}{T}- \ln \frac{T}{100 \GeV}\;,
\end{align}
where the numerical range comes from the uncertainty associated with the determination of the fluctuation determinant $\kappa = (10^{-4} - 10^{-1})$ as adopted by Dine {\it et. al.} \cite{Dine:1991ck} and is comparable to the uncertainty in the numerical lattice simulation of the sphalerons at the Standard Model electroweak (EW) 
crossover~\cite{DOnofrio:2014rug}.
In this work, we calculate $E_{\rm sph}(T)$ directly and test the BNPC by examining the quantity
\begin{align}\label{eq:PTsph}
PT_{sph}=\frac{E_{\rm sph}(T)}{T} - 7 \ln \frac{v(T)}{T} + \ln \frac{T}{100 \GeV}\;,
\end{align}
at the temperature of the EWPT, which is crucial for guaranteeing a successful baryon asymmetry generation.
We use the Standard Effective field theory (SMEFT) and the 
extensively studied singlet extended Standard Model (``xSM'') as two concrete examples and find that the BNPC condition can set a more rigorous bound on the new physics scale than the conventionally 
adopted SFOEWPT condition~\cite{Quiros:1999jp}: $\frac{v(T)}{T} \gtrsim 1 $.
We then check the scaling law, which states that the sphaleron energy at the temperature of phase transition ($E_{\rm sph}(T)$) 
and that at the zero temperature ($E_{\rm sph}$) obeys an approximate scaling relation~\cite{Braibant:1993is,Brihaye:1993ud}:
\begin{align}\label{Esph_scaling}
	E_{\rm sph}(T) \approx E_{{\rm sph}} \frac{v(T)}{v}\;,
\end{align}
where $v(T)$ and $v$ are the VEVs at the time of the phase transition and at zero temperature respectively.
Our analysis shows that the scaling law can be established when the strength of the phase transition increases to $PT_{sph}\sim \mathcal{O}(10^2)$, 
where the SFOEWPT points also meet the BNPC condition.
In this scenario, one usually has a smaller $\beta/H_n$ accompanied with a larger $\alpha$, and therefore a higher magnitude of the gravitational wave spectra, as shown in Section \ref{sec:GW}, which allows us to build a connection between the sphaleron energy and the gravitational wave spectra measurements.



\section{EWPT, Gravitational Waves and Sphalerons}
\subsection{The models} 
It is well known that the SM can not accommodate a first order EWPT and this has motivated a plethora of 
beyond the standard model scenarios with an extended Higgs sector. From an effective field theory point of view, a first order
EWPT can be realized by inclusion of higher dimensional operators, irrespective of a specific scenario.
Among the dimension-six operators of the SMEFT, the operator $O_6$ dominates the contribution to the Higgs potential.
Defining the SM Higgs doublet as  $H^{\text{T}} = (G^+, (v + h + i G^0)/\sqrt{2})$, we then have the following scalar potential:
\begin{align}\label{eq:V_phi}
	V(H) = -  m^2 (H^\dag H) +  \lambda \,(H^\dag H) ^2 + \frac{(H^\dag H)^3}{\Lambda^2} \;.
\end{align}
The presence of the last term allows the electroweak phase transition to be first order \cite{Grojean:2004xa,Delaunay:2007wb}, since taking $\mu^2 > 0$ and $\lambda < 0$ leads to a potential with a barrier between two minima.
The minimization condition and the Higgs mass definition lead to the relation
\begin{align}\label{eq:params}
&	m^2 =  \frac{1}{2} m_h^2 - \frac{3}{4} \frac{v^4}{\Lambda^2} \; , \quad \quad
	\lambda = \frac{m_h^2}{2v^2} - \frac{3}{2} \frac{v^2}{\Lambda^2} \;.
\end{align}
To study the EWPT, we need the finite temperature effective potential, which is given by\footnote{In the standard approach, one includes 
the tree level effective potential, the Coleman-Weinberg term~\cite{Coleman:1973jx} and its finite temperature counterpart~\cite{Quiros:1999jp}, together with the
daisy resummation~\cite{Parwani:1991gq,Gross:1980br}.
For the EWPT mainly driven by the cubic terms in the potential, and with a purpose of maintaining a gauge independent effective potential~\cite{Patel:2011th}, 
we use the gauge invariant high temperature expansion approximation~\cite{Profumo:2007wc,Profumo:2014opa,Kotwal:2016tex,Huang:2017jws,Alves:2018oct}.}
\begin{align}
V_T(h,T)=V(h) + \frac{1}{2} c_{hT} h^2\;,
\end{align}
where $c_{hT} = (4 y_t^2 +3g^2+g'^2+8\lambda)T^2/16$. 
The requirement of the EW minimum being the global one results in the condition $\Lambda\geq v^2/m_h$, and the EWPT can be first order when the potential barrier can be raised with $\Lambda<\sqrt{3}v^2/m_h$~\cite{Grojean:2004xa,Huang:2015tdv}.

Going beyond the framework of the SMEFT, a simplified benchmark model 
is the gauge singlet extension of the SM, known as the``xSM'', with the potential defined 
by~\cite{Profumo:2007wc,Profumo:2014opa,Huang:2017jws},
\begin{eqnarray}  \label{eq:v}
  V(H,S) &= -m^2 H^{\dagger} H + \lambda (H^{\dagger}H)^2
  + \frac{a_1}{2} H^{\dagger} H S  \nonumber \\
  &  + \frac{a_2}{2} H^{\dagger} H S^2 + \frac{b_2}{2} S^2 + \frac{b_3}{3} S^{3} + \frac{b_4}{4}S^4 \;,\nonumber
\end{eqnarray}
where $S=v_s + s$ is the real scalar gauge singlet.
The finite temperature potential is~\cite{Alves:2018jsw}:
\begin{eqnarray}  \label{eq:vt}
 && V[h,s,T] = - \frac{1}{2} [m^2 - \Pi_h[T]] h^2
  - \frac{1}{2} [-b_2 - \Pi_s[T]] s^2 \nonumber \\
  &&\hspace{0.7cm} + \frac{1}{4} \lambda h^4 + \frac{1}{4} a_1 h^2 s + \frac{1}{4} a_2 h^2 s^2 +
  \frac{b_3}{3} s^3 + \frac{b_4}{4} s^4, \quad \quad
\end{eqnarray}
where $\Pi_h(T)$ and $\Pi_s$ are the thermal masses of the fields,
\begin{eqnarray}
  &&  \Pi_h[T] = \left( \frac{2 m_W^2 + m_Z^2 + 2 m_t^2}{4 v^2} + \frac{\lambda}{2} + \frac{a_2}{24} \right) T^2, \nonumber \\
  &&  \Pi_s[T] = \left( \frac{a_2}{6} + \frac{b_4}{4} \right) T^2 \;.
\end{eqnarray}
The scalar cubic terms in Eq.~\ref{eq:vt} dominate the phase transition dynamics and can accommodate a first order EWPT 
after theoretical and experimental bounds on model parameters are taken into account.
Moreover, in this work, we focus on the one-step EWPT with the EW vacuum denoted by ($\equiv (v,v_s)$), though
two-step EWPT can also exist, which however is of negligible parameter space here~\cite{Alves:2018jsw}. 


\subsection{Gravitational Waves}
\label{sec:GW}
With the finite temperature effective potential given above, the Higgs VEV at finite temperature ($v(T)$) can be obtained.
Here and in the following sections we define the temperature of the EWPT as $T_\star\approx T_n$~\footnote{The approximation is justified for a EWPT 
without significant reheating~\cite{Caprini:2015zlo}}, with $T_n$ being the bubble nucleation temperature. 
The phase transition order parameter $v_n/T_n$ (the phase transition strength at the bubble nucleation temperature) and
the two crucial parameters for the GW spectrum from the EWPT are calculated using {\bf CosmoTransitions}
\cite{Wainwright:2011kj}. The first parameter crucial for the GW spectrum  is the ratio of released latent heat from the transition to the 
total radiation energy density~\cite{Caprini:2015zlo}
\begin{equation}\label{eqn:alpha}
\alpha= \left.\frac{1}{\rho_{R}}\left[-(V_{\rm EW}-V_f)+ T \left(\frac{dV_{\rm EW}}{dT} - \frac{dV_f}{dT}\right)\right]\right|_{T=T_*} ,
\end{equation}
where $V_f$ is the value of the potential at the metastable vacuum and $V_{\rm EW}$ is that in the EW vacuum. Another parameter $\beta/H_n$ serves as a time scale for the EWPT:
\begin{equation}\label{eqn:beta}
\frac{\beta}{H_n}= \left. \left[T \frac{d}{dT} \left(\frac{S_3(T)}{T} \right)\right]\right|_{T=T_*} ,
\end{equation}
where $H_n$ is the Hubble rate at $T_n$ and $S_3(T)$ the action for the $O(3)$ symmetric bounce action.

The dominant sources for GW production during the EWPT are the sound waves in the plasma~\cite{Hindmarsh:2013xza,Hindmarsh:2015qta} and
the magnetohydrodynamic turbulence (MHD)~\cite{Hindmarsh:2013xza,Hindmarsh:2015qta}\footnote{ 
We neglect here the contribution from the bubble wall collisions~\cite{Kosowsky:1991ua,Kosowsky:1992rz,Kosowsky:1992vn,Huber:2008hg,Jinno:2016vai,Jinno:2017fby}, as it is now generally believed to be negligible~\cite{Bodeker:2009qy}.}. 
To a good approximation, the total energy density of the gravitational waves 
in unit of the critical energy density of the universe is given by~\cite{Caprini:2015zlo}
\begin{equation}
\Omega_{\rm GW}(f) h^2  \approx\Omega_{\rm sw}(f) h^2  +\Omega_{\rm turb}(f) h^2.
\end{equation}
Due to its stochastic nature, this kind of gravitational waves can be searched for by cross-correlating 
outputs from two or more detectors, with the resulting signal-to-noise ratio(SNR) obtained as~\cite{Caprini:2015zlo}
\begin{eqnarray}\label{eq:snr}
  \text{SNR} = \sqrt{\mathcal{T} \int df
    \left[
      \frac{h^2 \Omega_{\text{GW}}(f)}{h^2 \Omega_{\text{exp}}(f)}
  \right]^2} ,
\end{eqnarray}
where $\mathcal{T}$ is the duration of the data in years and $\Omega_{\text{exp}}$ the power spectral density of the detector.

As shown in Appendix.~\ref{sec:GWfor}, the $\Omega_{\rm GW}(f) h^2$ is proportional to $\beta/H_n$, which has generally small values 
and is accompanied with a relatively large $\alpha$ for a strong phase transitions (a higher value of $v_n/T_n$)~\cite{Bian:2019zpn,Alves:2018jsw}.
Generally, one has $\alpha$ growing as $\Delta V\equiv (V_{\rm EW}-V_f)$ and $\beta/H_n$ scaling as 1/$\sqrt{\Delta V}$ for a given finite temperature potential~\cite{Grojean:2006bp}. 
This makes it possible to connect the sphaleron energy with the GW measurements since $E_{sph}(T_n)$ is proportional to $v_n/T_n$, and a large $v_n/T_n$  corresponds to a highly suppressed sphaleron rate inside the EW vacuum bubble as will be explored bellow.

\subsection{The Electroweak Sphaleron}

The electroweak sphaleron is a static but unstable solution to the classical equations of motion of the EW theory, which corresponds to a 
saddle-point configuration in the field space and sits at the top of the potential barrier between two topologically distinct vacua with adjacent values of 
the Chern-Simons number~\cite{Manton:1983nd,Klinkhamer:1984di}.
To calculate the energy of the sphaleron configuration, we adopt  
the spherically symmetric antasz since the ${\rm U(1)}_Y$ contribution is sufficiently small~\cite{Klinkhamer:1984di,Klinkhamer:1990fi,Kleihaus:1991ks,Kunz:1992uh}.
In the xSM, it follows that
\begin{align}\label{eq:E_sph}
	&E_{\rm sph}(T)
	= \frac{4 \pi \Omega[T]}{g} \int_{0}^{\infty} \! d \xi \, \Bigl[
	 4 \biggl( \frac{df}{d\xi} \biggr)^2	\nonumber\\
	& \qquad+ \frac{8}{\xi^2} f^2 \bigl( 1 - f \bigr)^2 + \frac{\xi^2v[T]^2}{2\Omega[T]^2} \biggl( \frac{dh}{d\xi} \biggr)^2\nonumber\\
	&\qquad
	+ \frac{v[T]^2}{\Omega[T]^2}  \bigl( 1-f \bigr)^2 h^2  +\frac{\xi^2v_s[T]^2}{2\Omega[T]^2}\biggl( \frac{dk}{d\xi} \biggr)^2	\nonumber\\ & \qquad
+\frac{\xi^2}{g^2\Omega[T]^4}(V_{eff}[h,k,T])
	\Bigr] ,
\end{align}
where $h,f,k$ are field configurations defined in Appendix~\ref{sec:sphcon}, $\xi=g_2 \Omega[T] r$, and $V_{eff}[h,k,T] = V[h,k,T] - \Delta[T]$, with $\Delta[T]$ being the cosmological constant energy density which can be regarded as the minimal value of the potential at temperature $T$~\cite{Ahriche:2007jp}. 
Here $4 \pi v[T] / g$ has the unit of energy and the integral gives a dimensionless number; $\Omega[T]$ can take any non-vanishing value of mass dimension one (for example $v[T]$,
$v_S[T]$ or $\sqrt{v[T]^2+v_S[T]^2}$);  $v[T],v_S[T]$ are the VEVs of $h,s$ at temperature $T$ and $v[T] = v$, $v_S[T]=v_S$ at $T = 0$.
When the singlet part is absent, the above $E_{{\rm sph}}(T)$ reduces to the form of the SMEFT case, with the potential:
\begin{align}
&V_{eff}[h,T] =V_T(h,T)-V_T(v(T),T)\;.
\end{align}
For more details on the field configurations and on the sphaleron solutions, see Appendix.~\ref{sec:sphcon}. 
The sphaleron energy at the bubble nucleation temperature $T_n$($E_{sph}(T_n)$) can be obtained after the parameters $v(T_n), v_s(T_n)$ and $T_n$ 
have been calculated through the EWPT analysis.

\section{Results}

\begin{figure}[!htp]
\begin{center}
\includegraphics[width=0.49\columnwidth]{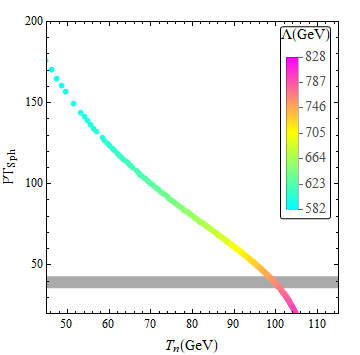}
\includegraphics[width=0.49\columnwidth]{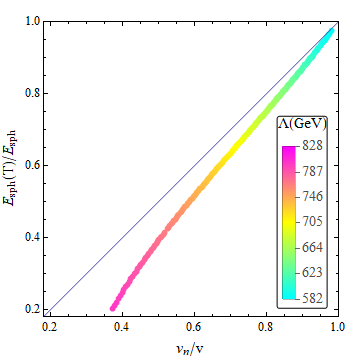}
\includegraphics[width=0.49\columnwidth]{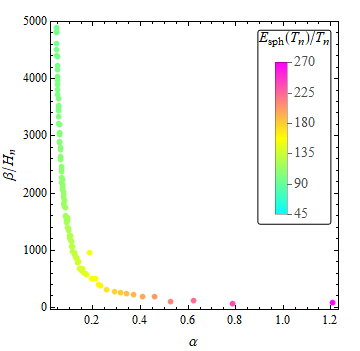}
\includegraphics[width=0.49\columnwidth]{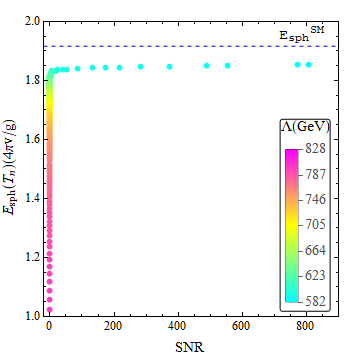}
\caption{In the top panels, we plot
$PT_{sph}$ versus $T_n$(left), and $E_{sph}(T_n)/E_{sph}$ versus $v_n/v$(right) when varying $\Lambda$ for the case of the SMEFT. The 
bottom panels show $\beta/H_n$ versus $\alpha$ by color-coding $E_{sph}(T_n)/T_n$(left), 
and the relation between the sphaleron energy $E_{sph}(T_n)$ and the SNR of the GW spectra(right).
}\label{fig:h6sphGW}
\end{center}
\end{figure}

In the top panel of Fig.~\ref{fig:h6sphGW}, we first present the relation between the BNPC and the new physics scale, where the horizontal 
shaded region represents the uncertainty of $\kappa$ in the sphaleron rate, i.e. the range $[35.9,42.8]$ in Eq.\ref{eq:Esph_bound} (this corresponds to $749<\Lambda<768$ GeV). When $PT_{sph}$ is above this region, the BNPC is satisfied, while $v_n/T_n\geq 1$ is obtained when $\Lambda<790$ GeV.   
The top-right panel demonstrates the scaling law,  where the deviation from the scaling law becomes smaller when the phase transition strength becomes 
larger obtained with a lower new physics scale $\Lambda$. 
We further present the relation between $E_{sph}(T_n)/T_n$ and the gravitational wave parameters $\alpha$ and $\beta/H_n$ in the bottom left panel. 
This demonstrates that a larger $\alpha$ (with a larger phase transition strength $v_n/T_n$) and a smaller $\beta/H_n$ generally lead to 
a larger $E_{sph}(T_n)/T_n$, for which the sphaleron rate is highly suppressed and washout of the baryon asymmetry can be avoided.
This correlation among these parameters is anticipated, as a larger $v_n$ can generally be obtained with a more significant supercooling.
The relation between the sphaleron energy and the SNR of the 
corresponding gravitational wave spectra from the EWPT is shown in the bottom-right plot, which indicates that 
the sphaleron can be probed by the gravitational wave detector (with a larger SNR) when $\Lambda$ is lower. 
The $E_{sph}(T_n)/T_n$ obtained for all EWPT points are smaller than the SM sphaleron energy at zero temperature.

\begin{figure}[!htp]
\begin{center}
\includegraphics[width=0.49\columnwidth]{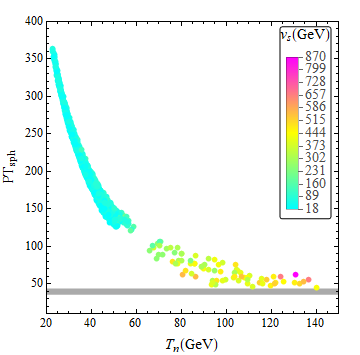}
\includegraphics[width=0.48\columnwidth]{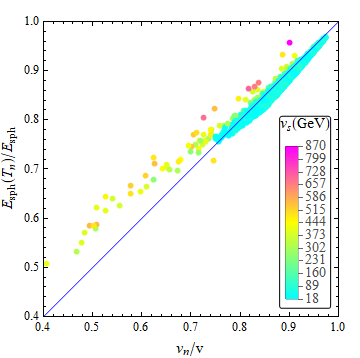}
\includegraphics[width=0.49\columnwidth]{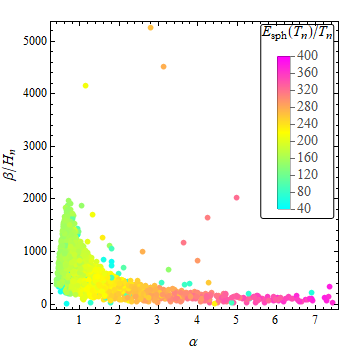}
\includegraphics[width=0.49\columnwidth]{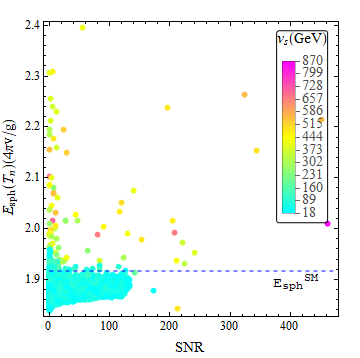}
\caption{As in Fig.~\ref{fig:h6sphGW}, we show similar plots with $v_n/T_n>1$ and SNR$>$10, where $v_s$ is the VEV of the singlet at zero 
temperature. }\label{fig:xsmspGW}
\end{center}
\end{figure}

Now we move to another scenario, where the new physics that is necessary for a first order EWPT cannot be
integrated out. We illustrate the situation with the xSM (see Ref.\cite{Damgaard:2015con,Huang:2015tdv} for the mismatch between the SMEFT and xSM model), 
and show in Fig.\ref{fig:xsmspGW} the points with SNR$>$10.
The top-left panel of Fig.\ref{fig:xsmspGW} shows that a lower $v_s$ generally leads to a lower $T_n$ and a larger $PT_{sph}$.
The top-right panel shows that the scaling law is better satisfied for small $v_s$.
We then examine the relation between $E_{sph}(T_n)/T_n$ and ($\alpha$, $\beta/H_n$) in the bottom-left panel, which indicates a similar behavior as the SMEFT case. The bottom-right panel shows that the sphaleron energy is concentrated at around 1.9 (in unit of $4\pi v/g$) for small $v_s$, where the GW signal can be probed by 
LISA~\cite{Caprini:2015zlo}. In the SNR calculation for the SFOEWPT points, we include the deficit found in the GW production from the sound waves~\cite{Cutting:2019zws}
\footnote{Note that the results of a recent numerical simulation by the same group shows that for strong EWPT (i.e., those with $\alpha \sim 1$), 
there is a significant deficit in the GW production from the sound waves~\cite{Cutting:2019zws}. This invalidates the
naive generalization of the GW formulae to arbitrary values of $(v_w, \alpha)$. For the xSM, this leads to a shrinking 
of the parameter space that gives detectable GWs while the main features of the resulting parameter space remains qualitatively unchanged(see~\cite{Alves:2019igs}
for a detailed study).
For the SMEFT, however, it leads to a more severe reduction of the parameter space, making it difficult to generate detectable
GWs.}. The $E_{sph}(T_n)/T_n$ for most SFOEWPT points are found to be smaller than the SM sphaleron energy at zero temperature.

\begin{figure}[!htp]
\begin{center}
\includegraphics[width=0.49\columnwidth]{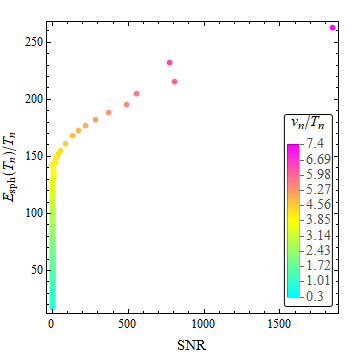}
\includegraphics[width=0.48\columnwidth]{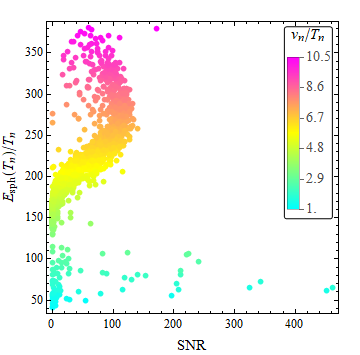}
\caption{We show the SNR of the GW spectra versus the $E_{sph}/T_n$ by color-coding $v_n/T_n$ for SMEFT (left panel) and xSM (right panel). }\label{fig:sphGW}
\end{center}
\end{figure}

We finally present Fig.~\ref{fig:sphGW}, which shows that a EWPT which is strong enough to produce an observable gravitational wave signal 
can effectively ``switch off'' the sphaleron rate after the transition as required by the EWBG.

%


\section{Conclusions and discussion}

We have calculated the energy of the electroweak sphaleron during the EWPT and revealed the relation between the BNPC and the new physics scale using the SMEFT and xSM as two concrete examples. The scaling law can be established approximately when there is a higher phase transition strength associated with a lower new physics scale. In this scenario, it is possible to use the GW detectors to access the sphaleron energy, thus providing an alternative way of probing the sphaleron 
in addition to the high energy colliders\footnote{Probing the (B+L)-violation process is of crucial importance to test the Sakharov conditions and the EWBG mechanism.
Previously, the conventional wisdom is that the (B+L)-violating process at zero temperature is difficult to probe at current and near future high energy colliders 
as the sphaleron induced (B+L)-violating process is highly rare~\cite{Rubakov:1996vz,Bezrukov:2003er,Bezrukov:2003qm,Ringwald:2002sw,Ringwald:2003ns}. Very recently, the Bloch wave approach was proposed such that there is a chance to observe
a ($B + L$)-violating event at the Large
Hadron Collider (LHC)~\cite{Ellis:2016ast,Qiu:2018wfb,Tye:2017hfv,Ellis:2016dgb,Tye:2015tva}.}.
Different from other sources of the gravitational waves such as cosmic strings, domain walls, and primordial black hole, 
the GWs from SFOEWPT can also be tested through future colliders, since the SFOEWPT usually is accompanied by a deviation 
of the triple and quartic Higgs couplings in the Higgs potential
\footnote{The sensitivities of future colliders in the SFOEWPT parameter space (where one can have gravitational wave signal) can be accessed with the measurements of the two couplings ($\lambda_{hhh}$ and $\lambda_{hhhh}$) at future $e^+e^-$ colliders and the HL-LHC~\cite{DiVita:2017vrr,Liu:2018peg}.}.
Thus the relation between the sphaleron energy and the phase transition strength studied here makes it possible to probe the sphaleron through 
both gravitational wave and collider measurements, in a complementary role.

At last, we note that our findings in this work rely only mildly on the numerical uncertainty in Eq.~\ref{eq:Esph_bound}.
Settling down the numerical range for a specific particle physics model would require a lattice simulation of the sphaleron rate~\cite{Moore:1998swa}\footnote{
The starting point of the lattice simulation of the sphaleron rate and EWPT is the dimensional reduction. 
For the SMEFT theory with dimensional six operators,  the 3d EFT obtained after dimensional reduction is not super-renormalizable, and the lattice-continuum relations receive corrections at all orders in perturbation theory. To study the system at small lattice spacings, the difficulty is the determination of the non-perturbative relation between physical inputs and quantities measured on the lattice~\cite{Andersen:2017ika,Gorda:2018hvi,Kainulainen:2019kyp,Farakos:1994kx,Kajantie:1995dw}). 
}. 

\section*{Acknowledgments}
The work of L.B. is supported by the National Natural Science Foundation of China under grant No.11605016 and No.11647307. 
H.G. is partially supported by the U.S. Department of Energy grant DE-SC0009956. We thank F.R. Klinkhamer, Mark Hindmarsh, Salah Nasri, Guy D. Moore, Mikko Laine, Lauri Niemi, Kari Rummukainen, Lian-Tao Wang, Koichi Funakubo, and Heng-Tong Ding for helpful communications and discussions.

\appendix
\section{The GW Energy Density Spectra}
\label{sec:GWfor}

It is realized in recent years that the long lasting sound waves in the plasma during and after the phase transition
constitutes the dominant GW source~\cite{Hindmarsh:2013xza,Hindmarsh:2015qta}. The energy density spectrum from this source is obtained by large scale
numerical simulations, based on the scalar field and fluid model, for weak phase transitions corresponding to small 
values of $v_w$ and $\alpha$. It is well fitted by~\cite{Hindmarsh:2015qta}
\begin{eqnarray}
  &&\Omega_{\textrm{sw}}h^{2}=2.65\times10^{-6}\left( \frac{H_{\ast}}{\beta}\right) \left(\frac{\kappa_{v} \alpha}{1+\alpha} \right)^{2}
\nonumber\\
&&\left( \frac{100}{g_{\ast}}\right)^{1/3}
 v_{w} \left(\frac{f}{f_{\text{sw}}} \right)^{3} \left( \frac{7}{4+3(f/f_{\textrm{sw}})^{2}} \right) ^{7/2} \ ,
\label{equ:soundwaves}
\end{eqnarray}
where $H_{\ast}$ is the Hubble rate at $T_{\ast}$ when the phase transition finished, which is only slightly different from $T_n$;
$v_w$ is the bubble wall velocity and is chosen so that a non-relativistic relative velocity in the bubble wall frame can be obtained 
to make sure the slower baryon generation process is feasible~\cite{No:2011fi,Alves:2018jsw,Alves:2019igs};
$g_{\ast}$ is the relativistic degrees of freedom. 
The factor $\kappa_v$ denotes the fraction of released energy density that is transferred into the kinetic energy of the plasma, 
which can be calculated given inputs of $v_w$ and $\alpha$ from a hydrodynamic analysis~\cite{Espinosa:2010hh}.
Moreover $f_{\text{sw}}$ is the peak frequency:
 \begin{equation}
f_{\textrm{sw}}=1.9\times10^{-5}\frac{1}{v_{w}}\left(\frac{\beta}{H_{\ast}} \right) \left( \frac{T_{\ast}}{100\textrm{GeV}} \right) \left( \frac{g_{\ast}}{100}\right)^{1/6} \textrm{Hz} .
\end{equation}
We note that the above formulae is limited to relatively small values of $v_w$ and $\alpha$. Recent numerical simulations exploring larger values of $\alpha$
shows a deficit in the GW production~\cite{Cutting:2019zws}(see~\cite{Alves:2019igs} for a more detailed discussion on the implications of this effect).

There is also a small fraction of energy going to the MHD, with a result that can be fitted by~\cite{Caprini:2009yp,Binetruy:2012ze}
\begin{eqnarray}
  &&\Omega_{\textrm{turb}}h^{2}=3.35\times10^{-4}\left( \frac{H_{\ast}}{\beta}\right) \left(\frac{\kappa_{\text{turb}}
\alpha}{1+\alpha} \right)^{3/2} \nonumber\\
&&\left( \frac{100}{g_{\ast}}\right)^{1/3}
 v_{w}  \frac{(f/f_{\textrm{turb}})^{3}}{[1+(f/f_{\textrm{turb}})]^{11/3}(1+8\pi f/h_{\ast})} \;, \quad
\label{eq:mhd}
\end{eqnarray}
where $\kappa_{\text{turb}}$ is the fraction of the energy transferred to the MHD turbulence and is
approximately given by $\kappa_{\text{turb}} \approx (0.05 \sim 0.1) \kappa_{v}$. We take here $\epsilon\approx 0.1 $.
Finally $f_{\text{turb}}$ is the peak frequency this spectrum:
\begin{equation}
f_{\textrm{turb}}=2.7\times10^{-5}\frac{1}{v_{w}}\left(\frac{\beta}{H_{\ast}} \right) \left( \frac{T_{\ast}}{100\textrm{GeV}} \right) \left( \frac{g_{\ast}}{100}\right)^{1/6} \textrm{Hz} .
\end{equation}

\begin{figure*}[t]
\centering
\includegraphics[width=\columnwidth]{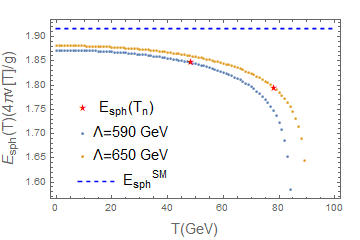} 
\includegraphics[width=\columnwidth]{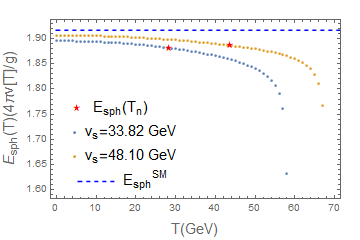}
\caption{Left: $E_{sph}(T_n)$ versus $T$ for the SMEFT,  SNR$=175 (0.002)$ and $v_n/T_n=4.8 (2.6)$ for $\Lambda=590(650)$ GeV;Right: xSM scenario for $v_s=33.82 (48.10)$ GeV, with $v_n/T_n=8.3 (5.1)$ and $SNR=98 (11)$. }\label{fig:h6spGW}
\end{figure*}

\section{Sphaleron configurations}\label{sec:sphcon}
For the xSM model, we consider the following sphaleron field ansatz~\cite{Fuyuto:2014yia}:
\begin{align}
 A_i(\mu,r,\theta,\phi)
 &=-\frac{i}{g}f(r)\partial_iU(\mu,\theta,\phi)U^{-1}(\mu,\theta,\phi),\label{antazA}\\
H(\mu,r,\theta,\phi)
&=\frac{v(T)}{\sqrt{2}}\bigg[(1-h(r))
	\bigg(
	\begin{array}{c}
	0 \\
	e^{-i\mu}\cos\mu
	\end{array}
	\bigg)\nonumber\\
&+h(r)U(\mu,\theta,\phi)
	\bigg(
	\begin{array}{c}
	0 \\
	1
	\end{array}
	\bigg)\bigg], \label{antazH}\\
S(\mu, r, \theta, \phi) &=v_S(T)k(r),
\end{align}
where $A_i$ are SU(2) gauge fields, and the matrix $U$ is defined as
\begin{eqnarray}
&&U(\mu,\theta,\phi)=\nonumber\\
	&&\bigg(
	\begin{array}{cc}
	e^{i\mu}(c_\mu-i s_\mu c_\theta) & e^{i\phi}s_\mu s_\theta \\
	-e^{-i\phi}s_\mu s_\theta & e^{-i\mu}(c_\mu+i s_\mu c_\theta)
	\end{array}
	\bigg),
\end{eqnarray}
where the $s_{\mu(\theta)}=\sin\mu(\theta)$ and $c_{\mu(\theta)}=\cos\mu(\theta)$. The sphaleron energy is obtained for $\mu=\pi/2$~\cite{Klinkhamer:1984di}.
From Eq.~(\ref{eq:E_sph}), the equations of motion can be found:
\begin{align}
& \frac{d^2f}{d\xi^2}
= \frac{2}{\xi^2}f(1-f)(1-2f)
 -\frac{v[T]^2h^2}{4\Omega[T]^2}(1-f),\\
& \frac{d}{d\xi}\left(\xi^2\frac{dh}{d\xi}\right)
= 2h(1-f)^2
+\frac{\xi^2}{g^2}\frac{1}{v[T]^2\Omega[T]^2}\frac{\partial V_{\rm eff}(h,k,T)}{\partial h}, \\
&\frac{d}{d\xi}\left(\xi^2\frac{dk}{d\xi}\right)
= \frac{\xi^2}{g^2}\frac{1}{v_S[T]^2\Omega[T]^2}\frac{\partial V_{\rm eff}(h,k,T)}{\partial k}.
\end{align}
The sphaleron solutions can be obtained with the following boundary conditions,
\begin{eqnarray}
&&\lim_{\xi\to0} f(\xi) = 0, \lim_{\xi\to0} h(\xi) = 0,\quad \lim_{\xi\to0} k'(\xi) = 0,  \nonumber\\
&&\lim_{\xi\to\infty} f(\xi) = 1, \lim_{\xi\to\infty} h(\xi) = 1, \lim_{\xi\to\infty} k(\xi) = 1.
\end{eqnarray}

For the SMEFT, the sphaleron solutions can be obtained from:
\begin{align}
&\frac{d^2f}{d\xi^2}
= \frac{2}{\xi^2}f(1-f)(1-2f)-\frac{1}{4}h^2(1-f),\nonumber\\
&\frac{d}{d\xi}\left(\xi^2\frac{dh}{d\xi}\right)
= 2h(1-f)^2  + \frac{\xi^2}{g^2}\frac{1}{v[T]^4} \frac{\partial V_{\rm eff}(h,T)}{\partial h} ,\;
\end{align}
with boundary conditions given by,
\begin{eqnarray}
&&\lim_{\xi\to0} f(\xi) = 0, \quad \ \lim_{\xi\to0} h(\xi) = 0, \nonumber\\
&&\lim_{\xi\to\infty} f(\xi) = 1, \quad \lim_{\xi\to\infty} h(\xi) = 1\;.
\end{eqnarray}
 We implement the relaxation method as documented in {\it Numerical Recipes}~\cite{Press:2007:NRE:1403886} to solve the above ordinary differential equations
numerically.

Here we present Fig.~\ref{fig:h6spGW} to illustrate that the sphaleron energy at the temperature of the phase transition 
approaches the corresponding value at the zero temperature when one has a lower $\Lambda$ for SMEFT (or $v_s$ in xSM scenario), 
accompanied with a higher value of the phase transition strength and a larger SNR of the GW spectra. 
The value $E_{sph}(T)$ grows as the temperature of the universe
decreases, and thus we have a increasingly suppressed sphaleron rate inside the electroweak bubbles after the EWPT.

\bibliographystyle{utphys}
\bibliography{mybib}

\end{document}